\documentstyle[12pt,epsf]{article}
\def\lamF{\lambda_F}
\def\lamG{\lambda_G}
\def\mrec{m_{rec}}
\def\BAR{\overline}
\def\pra{|{\rm PRA}|}
\def\fx{F_x}
\def\gx{G_x}
\def\uglu{\hskip 0pt plus 1fil minus 1fil}
\def\uglux{\hskip 0pt plus .75fil minus .75fil}
\def\slashed#1{\setbox200=\hbox{$ #1 $}
    \hbox{\box200 \hskip -\wd200 \hbox to \wd200 {\uglu $/$ \uglux}}}
\def\slq{\slashed q}

\begin{document} 
\begin{center}

{\large\bf Desperately Seeking Non-Standard Phases via \\ 
Direct CP Violation in $b\to sg^\ast$ Process}\\
%% {\small \bf version 16.0}\\
\bigskip

D. Atwood$^a$ and A. Soni$^b$ \\
\bigskip

$a$) Theory Group, CEBAF, Newport News, VA\ \ 23606 \\
$b$) Theory Group, Brookhaven National Laboratory, Upton, NY\ \ 11973 
\bigskip\bigskip
\end{center}

\begin{quotation}
{\bf Abstract:}
Attributing the recent CLEO discovery of $B \to \eta' + X_s$ to
originate (primarily) from the fragmentation of an off-shell gluon ($g^*$) via
$b \to s + g^*$, $g^* \to g + \eta'$, we emphasize that many
such states ($X_g$) should materialize. 
Indeed the hadronic fragments
($X_g$) of $g^*$ states are closely related to those seen in
$\psi \to \gamma (\phi, \omega) + X_g$. 
A particular final state of considerable interest is $X_g=K^+K^-$.
Signals from such states in $B$ decays can be combined to provide a very
sensitive search for CP violating phase(s) from non-standard physics. 
The method should work even if the contribution of these source(s)
to the rates is rather small ($\sim10\%$) 
to the point that a comparison between theory and experiment
may find it extremely difficult to reveal the presence of such a new
physics.
\end{quotation}

\vspace*{.25 in}

%%%%%%%%%%%%%%%%%%%%%%%%%%%%%%%%%%%%%%%%%%%%%%%%%%%%%%%%%%%%%%%%%%%%%
%%%%%%%%%%%%%%%%%%%%%%%%%%%%%%%%%%%%%%%%%%%%%%%%%%%%%%%%%%%%%%%%%%%%%
%%%%%%%%%%%%%%%%%%%%%%%%%%%%%%%%%%%%%%%%%%%%%%%%%%%%%%%%%%%%%%%%%%%%%
%
%
%     main text
%
%%%%%%%%%%%%%%%%%%%%%%%%%%%%%%%%%%%%%%%%%%%%%%%%%%%%%%%%%%%%%%%%%%%%%
%%%%%%%%%%%%%%%%%%%%%%%%%%%%%%%%%%%%%%%%%%%%%%%%%%%%%%%%%%%%%%%%%%%%%
%%%%%%%%%%%%%%%%%%%%%%%%%%%%%%%%%%%%%%%%%%%%%%%%%%%%%%%%%%%%%%%%%%%%%

Rather compelling theoretical arguments suggest the existence of CP
violation phase(s) originating from physics beyond the Standard Model
(SM). For one thing it is extremely difficult to understand
baryogenesis in the SM whereas it becomes quite plausible in many of
its extensions\cite{cohen}. Of course existence of the three quark families
endows the SM its CKM phase\cite{KMref}  and,  as a rule, it requires some
degree of contrived physics 
to get rid of this phase. But, new physics, whether it involves extra
Higgs, fermions and/or gauge bosons, rather naturally entails
additional CP violating phases. Therefore, on quite general grounds, we
should expect non-standard phase(s) in addition to the CKM phase of the
SM\null. It is clearly very important then, to search for these
phase(s).

In $B$-physics significant CP violating asymmetries are expected to occur
in many processes. In $b\to s$ transitions, however, such asymmetries will
be small  as in the standard Wolfenstein representation\cite{wolf,quinn} 
this phase is manifestly small ($0(\eta\lambda^2)$).
Thus looking for CP violation in $b\to s$ transitions can be a powerful method
to look for CP violation from non-standard phases(s)\cite{gronau}. 
%
%
%
%%%%%%%%%%%%%%%%%%%>>>>>>>>>>>DA
%
%  In $B$-physics, while significant (CP violating) asymmetries
%  are expected to occur in many processes, in $b\to s$ transitions, in the
%  standard Wolfenstein representation\cite{wolf,quinn} 
%  the phase is expected to be very
%  small ($0(\eta\lambda^2)$) and the SM driven effects in direct CP
%  violation tend to be rather suppressed. Thus $b\to s$ transitions can be
%  a powerful handle for searching for non-standard phase(s)\cite{gronau}. 
%
%%%%%%%%%%%%%%%%%%%
%
%
%
In particular, since
$b\to sg^\ast$ has a rather hefty branching ratio\cite{hss}
(1--2\%) it would be very helpful if this class of modes could be used
towards that goal. The $g^\ast$ in this decay carries only a few GeV
energy and it is 
very
difficult to perform a completely inclusive
search for $b\to sg^\ast$. Fortunately, recent experiments at CLEO have
provided an extremely important clue that enables us now to hunt for
fragments of $b\to sg^\ast$.

Perhaps one of the most important recent development in $B$-physics is
the CLEO discovery\cite{CLEOeta}:

\begin{equation}
Br(B\to \eta^\prime +X_s ; 2.2\le E_{\eta^\prime} \le 2.7\,{\rm GeV}) =
(7.5 \pm 1.5\pm 1.1) \times 10^{-4} \label{eqone}
\end{equation}

%%>>>DA
%    This finds a rather 
%%

\noindent  It  has been suggested that this finds a  natural explanation 
in the SM as originating from $b\to sg^\ast$ via the fragmentation:
$g^\ast\to g\eta^\prime$\cite{ASref}.
Of course, as always, quantitative calculations of purely hadronic modes
are extremely uncertain\cite{ASref_a}  but it seems fairly safe
to assume that an appreciable fraction of the observed signal has this
SM origin. It thus becomes important to ask what other distinctive
states can $g^\ast$ fragment into in an analogous fashion.
From an experimental point of view this question is
especially significant given the fact that the $\eta^\prime$ detection
%AS1
efficiency is (at least for now) small, only about 5\%\cite{CLEOeta}.

Theoretically, on very general grounds, one expects that $g^\ast\to gX_g$
should lead to those states $X_g$ that have the quantum numbers of $I(J^{PC}) =
0(0^{-+})$, $0(0^{++}), 0(2^{++})$ etc. In addition to these single particle
states, from the experimental perspective there are also interesting continuum
of states such as $\pi\pi$, $K\BAR K$, $\pi\pi K\BAR K$\dots Indeed there is a
close correspondence between these states $X_g$ (expected in $g^\ast\to gX_g$)
and those observed in $\psi\to\gamma (\phi,w)+X_g$. In $\psi$ decays these
states result from fusion of two gluons. Crossing symmetry provides a close
link between $\psi\to\gamma(w,\phi)X_g$ and $g^\ast\to gX_g$. Thus in $B\to
X_sX_g$ the experimentalists should be seeking states $X_g$ that have been seen
in $\psi\to\gamma(\phi,w) X_g$. Explicit examples of interesting states
are\cite{PDB,glueballs}:

\begin{eqnarray}
0^{-+} & : & \eta(958), \quad \eta(1440), \quad \eta(1295), \quad
\eta(1760)\dots \nonumber \\
0^{++} & : & f_0(980), \quad f_0(1370), \quad f_0(1500)\dots \nonumber
\\
2^{++} & : & f_2(1270), \quad f^\prime_2(1525), \quad f_2(2010), \quad
f_2(2300), \quad f_2(2340)\dots \nonumber \\
{\rm Continuum} & : & \pi\pi, \quad K\BAR K, \quad K\bar K \pi,
\quad 4\pi, \quad 4K, \quad \pi\pi K\BAR K\dots
\end{eqnarray}

Fig.~1a shows the SM fragmentation process leading to these states. This SM
amplitude (Fig.~1a), as explained above, has a vanishingly small CP-odd (CKM)
phase in the Wolfenstein convention. For simplicity, in this work, we will set
it to zero. Fig.~1b represents some non-standard physics scenario which
contributes to the same final state but has a CP-odd (non-standard) phases
$\lamF$ in the chromo-electric moment and $\lamG$ in the chromo-magnetic
moment.  Although Fig.~1a does not have a CP-odd phase it does possess a
CP-even phase ($\delta_{st}$) originating primarily from the $c\BAR c$ cut in
the Feynman amplitude\cite{bander}. The crucial point is that to this order in
perturbation theory $\delta_{st}$ is only a function of $q^2$ ($q$ is the
4-momentum of $g^\ast$) and does not depend on the state $X_g$. Thus the same
$\delta_{st}$ tends to show up in all the states $X_g$ that result from the
fragmentation of $g^\ast$. Therefore the CP asymmetries can be combined to
significantly improve our chances of capturing such a phase 1) due to the
vastly improved statistics and 2) due to the fact that many of the states can
have improved detection efficiency compared to the 5\% of the
$\eta^\prime$\cite{CLEOeta}.

Of course perturbative arguments should be taken with reservation as
they are bound to be corrections. But they are some reasons to think
that corrections may not be very big. For one thing $\delta_{st}$ is a
ratio of the imaginary to the real part of the Feynman amplitude and
therefore corrections ought to be less than on each of these
amplitudes. Also in QCD, the $0^{-+}$ (e.g.\ $\eta^\prime(958)$)
and $0^{++}$  (e.g.\ $f_0(980)$)\cite{PDB,cornwall} states have
considerable similarity. In this regard notice in particular that 
$Br(\psi\to\phi \eta^\prime(958))  =  (3.3\pm.4) \times 10^{-4}$,
$Br(\psi\to\phi f_0(980)) =  (3.2\pm.9) \times 10^{-4}$; also
$Br(\psi\to \omega\eta^\prime(958))  =  (1.67\pm .25) \times 10^{-4}$,
and
$Br(\psi\to w f_0(980)) =  (1.4\pm .5) \times 10^{-4}$.
Furthermore, for $\pi\pi$ and $K\BAR K$, to the extent that SU(3) is a
good symmetry, $\delta_{st}$ in $\pi\pi$ must equal $\delta_{st}$ in
$K\BAR K$ for a fixed invariant mass of the pair. Finally, in this
regard the most important point is that the theory need not be used to
predict the actual magnitudes of the asymmetries. 
Based purely on the quantum numbers (i.e.\ CP-odd or CP-even) of the states,
theory provides a reliable basis for combining the asymmetries observed in
individual channels no matter what the actual magnitudes of the asymmetries
might be. Thus theoretical considerations are allowing one to combine the
statistics to extend  the reach of the experiments without sacrificing
the quantitative detail to the altar of our theoretical inability to
accurately predict $\delta_{st}$ via perturbation theory.

The formalism is very simple. We begin with writing a general form for
the $bsg$ vertex\cite{hel_note}:

\begin{equation}
\Lambda^{bsg}_\mu = \frac{V_tG_F}{\sqrt{2}}\  
\BAR s_i  T^a_{ij}\left[ -i F(q^2)
(q^2 \gamma_\mu - q_\mu \slq) L + \frac{g_s}{2\pi^2} m_b q_\mu
\epsilon_\nu\sigma^{\mu\nu} G(q^2) R\right] b_j 
\end{equation}
where  $V_t  =  V_{tb}V^*_{ts}$,
\begin{eqnarray}
F(q^2) = e^{i\delta_{st}} F_{SM} + e^{i\lamF} \fx \ \ ; \ \ \ \ 
G(q^2)  =  G_{SM} + e^{i\lamG} \gx 
\end{eqnarray}
and $i$, $j$ and $a$ are suitable color indices. Here the $F$ and $G$ are
chromo-electric and  chromo-magnetic\cite{kagan,hou_note} form factors. The
strong phase, $\delta_{st}$ is generated by the imaginary part resulting
from the $c\bar c$ cut when  $q^2>4m_c^2$. 
%
% thus $F^R_{SM}=\cos\delta_{st} |F_{SM}|$ and
% $F^I_{SM}=\sin\delta_{st} |F_{SM}|$. 
%
The subscripts SM indicates a SM origin and $x$ indicates beyond the SM\null.
The differential decay rate takes the form:

\begin{equation}
\frac{d\Gamma}{dsdt} = F^2\Gamma_1 + G^2\Gamma_2 +2FG\Gamma_3 
\end{equation}

\noindent The CP-even and CP-odd contribution to the decay rates for
$B\to X_sX_g$ and $\BAR B\to \BAR X_s\BAR X_g$ can be written as 

\begin{eqnarray}
\Gamma_s & = & \frac{1}{2}\, \frac{d(\Gamma+\BAR \Gamma)}{dsdt} =
(F^2_{SM} + F^2_x +2F_{SM}\fx \cos \lamF) \Gamma_1 + \nonumber \\
& & \quad ( (G_{SM})^2 
+ G^2_x +2G_{SM}\gx \cos\lamG) \Gamma_2 + 2(F_{SM}G_{SM} \nonumber \\
& & \quad +\fx G_{SM}\cos \lamF + F_{SM}\gx  \cos\lamG
+\fx \gx \cos(\lamF-\lamG)) \Gamma_3  \nonumber\\
\Gamma_A & = & \frac{1}{2}\, \frac{d(\Gamma-\BAR\Gamma)}{dsdt} = -2
\sin\delta_{st}  F_{SM} (\fx \Gamma_1\sin\lamF+\gx \Gamma_3\sin\lamG) 
\end{eqnarray}

\noindent
Here $s=(p_b-p_s)^2$ and $t=(p_s+p_g)^2$; $p_{b,s}$ 
is the momentum of the $b,s$-quark and $p_g$ the momentum of the gluon.

The important CP violation observable is the partial rate asymmetry
(PRA) which in its differential form is obtained by the ratio:

\begin{equation}
{\cal A}_{PRA}(s,t) = \Gamma_A/\Gamma_s 
\end{equation}

\noindent In these equations $\Gamma_1$, $\Gamma_2$ and $\Gamma_3$ are
functions of $s$, $t$ and the masses $m_{X_g}$, $m_b$.

We can then take these expressions for the $b$-quark decay and put them
in the $B$-meson system with Fermi-motion and obtain the necessary
distributions for the rate and PRA as a function of $\mrec$\cite{ASref} where
$\mrec^2=(p_B-p_{X_g})^2$ is the mass recoiling against $X_g$.
Fig. 2 shows the $|{\cal A}_{PRA}|$ as a function of $\mrec$
%AS1
where, for concreteness, we have used $\sin\lamF=\sin\lamG=1$. Fig. 2a is
with the assumption that non-standard physics (NSP) contributes 10\% to the
total rate for each channel. Fig. 2b is similar except with the assumption
that NSP contributes 50\% to the rate. The PRA tends to scale with the amplitude
of NSP  so the  PRA for the second case is larger by factors of about 2.
For simplicity we also show separately the case where the NSP is all due
to the chromo-electric (CE) form factor ($\fx $) versus when it is all due
to a chromo-magnetic (CM) form factor ($\gx $). Notice that the CE case tends
to generate 50-100\% larger asymmetries compared to the CM case. 
%AS2
Fig. 2c compares the $|PRA|$ for all the three $J^{PC}$ of interest
assuming $G_x=0$ and the NSP, due to the CE form-factor alone, 
contributes 10\% to the rate. 
The PRA
shows some dependence on the $J^{PC}$ of the state; the asymmetries for the
$2^{++}$ states are thus somewhat bigger than for the $0^{++}$ which in turn
are bigger than for the $0^{-+}$. For a fixed $J^{PC}$ the PRA tends to
slowly increase with the mass of the state. The crucial point to note is
that the PRA for the $0^{++}$ and  the $2^{++}$ will, in general, 
all have the same sign and that for the $0^{-+}$ will be
opposite. The asymmetries in different channels can be combined after taking
this sign change into account.

      An important background originates from processes such as 
$B\to D(D_s)+X_g+X$. Most of this background gives $\mrec\geq 2GeV$ for various
$X_g$ states of interest. In the signal region, i.e. $\mrec\leq 2GeV$, the PRA
tends to show only little dependence on $\mrec$ (see fig.~2).
%AS2
For the sake of completion, in Fig.3, we also show the $\mrec$
distribution for the three $J^{PC}$.

We now briefly discuss some of the interesting final states. For $f(980)$ the
$\pi\pi$ and $K\BAR K$ are nice modes with branching ratios $\sim 78\%$ and
$\sim 20\%$ respectively\cite{PDB}. For $f_0(1500)$ the modes of interest
include $\eta\eta$ and $4\pi$. $f_2^\prime(1525)$ has a notably large ($\sim
90\%$) branching ratio to $K\BAR K$ with about $10\%$ branching ratio to
$\eta\eta$\cite{PDB}. For $0^{-+}$ states (e.g. $\eta(1440)$) the prominent
modes include three body modes such as $K\BAR K\pi$ and $\eta \pi\pi$ as well
as $a_0(980)\pi$ and $4\pi$\cite{PDB,eta*note}. $f_2(2300)$ and $f_2(2340)$
can also decay to $\phi\phi$ states\cite{PDB}.

%>>>>>>>>>>>>>>>>>here
Recall that the overall signature for these modes (i.e. $B\to X_s+X_g$) is that
%AS1
the $X_s$ must hadronize as a $K+n\pi$\cite{CLEOeta}.
Application of CP tests
require that $B$ and $\BAR B$ be distinguished. A number of strategies can be
used here. For one thing, the ``other $B$''. can be tagged e.g.\ through its
semi-leptonic decay modes. Typically this can have a tagging efficiency of
about 20\% (including both $e^\pm$ and $\mu^\pm$ final states). Also
``self-tagging'' can be used. This means that to a high degree of accuracy the
net charge carried by the kaons will distinguish $b$ from $\bar b$. Since the
$s$ (or $\BAR s$) quarks in the $b\to s$ or $\BAR b\to \BAR s$ transition will
give a $K^-$ (or $K^+$) about half the time, this method will have an
efficiency of about 50\%. Thus the total tagging efficiency will be about 60\%.

We want to emphasize that non-resonant, continuum originating from the
fragmentation of the $g^*$ can also be very useful. Perhaps the best example
here is the case of $K\BAR K$. First of all notice from $\psi$ decays that
$Br(\psi\to\phi K\BAR K)/Br(\psi\to \phi\eta^\prime)\sim 4$. Thus the $K\BAR K$
final state could well 
appear in appreciable fraction as $X_g$ (i.e.\ $g^*\to g
K\BAR K$) in $B$ decays as well. Such a final state should be rather
distinctive as it will have three kaons in the final state.

Since a $K\BAR K$ final state cannot be in a CP-odd configuration, resonant or
non-resonant, it will only be in one of the CP-even states with ``natural''
$J^{PC}$, i.e.\ $0^{++}$, $1^{--}$, $2^{++}$ etc.\cite{vector_note}. 
Thus statistics for the PRA in $K\BAR K$ states, whether produced through
a resonance or the 
%AS1
continuum, may be combined. Since the PRA varies little with the invariant
mass  or the total spin of the $K\BAR K$ system (see fig. 2), the
breakdown of the total $K\BAR K$ sample into various resonance and continuum
states will not greatly effect the resultant PRA. In the continuum state the
invariant mass of the $K\BAR K$ will play the role of $m_{X_g}$. 
Clearly similar comments apply to $\pi\pi$ states.

Let us now try to estimate the reach of the proposed experiments. For this
purpose we will take the new physics to contribute $\sim 10 \%$ to the rate for
each of these states. Due to its distinctiveness let us first consider
$X_g=K^+K^-$. The expected PRA (see fig.~2a) ranges between $13\%$ and 
$18\%$; we will take it to be $15\%$. We further estimate the detection 
($\epsilon_d$) and tagging ($\epsilon_b$) efficiencies to be $60\%$ each.
Although $\psi$ decays suggest that $g^*\to g K^+K^-$ could even be appreciably
bigger than than $g^*\to g\eta^\prime$, for our illustrative purpose we assume
they are the same. Thus the number ($N^{3\sigma}$) of $B$ ($\BAR B$) 
(i.e.\ $N^{3\sigma}$  is the total number of $B$ plus the total number of
$\BAR B$) mesons  needed to see a signal with $3\sigma$ significance in the
$K^+K^-X_s$ channel is:
\begin{eqnarray}
N^{3\sigma}={9\over Br {\cal A}_{PRA}^2\epsilon_d^{K^+K^-}\epsilon_b}
\approx 1.5\times 10^6
\end{eqnarray}
This means that the existing data sample at CLEO ($\sim 2\times 10^{6}$)
may already be able to provide a useful indication for non-standard physics.

To gain a second perspective let us consider now the case of various individual
resonances, in particular those with $J^{PC}=0^{-+}$, $0^{++}$ and $2^{++}$.
The PRA in each channel can be $8-18\%$ (see fig. 2a); for concreteness we
take PRA$\sim 10\%$. For the purpose of making an estimate let us further
assume that by combining many of the modes listed above the effective Br
is roughly three times that of the $\eta^\prime$. Also in this example we
take the effective detection efficiency ($\epsilon_d$) to be 20\% and the
tagging efficiency  ($\epsilon_b$) to be 60\%.
%AS1
The required number of $B$($\BAR B$) to see a $3-\sigma$ signal is then 
given by
\begin{eqnarray}
N^{3\sigma}={9\over Br^{eff} {\cal A}_{PRA}^2\epsilon_d\epsilon_b}
\approx 4\times 10^6
\end{eqnarray}
%
%
%

%AS1
Thus, from eqs.~(8--9), we see that with a sample of several million $B\BAR
B$ pairs the presence of a CP violating phase from physics beyond the SM
could be detected even if the contribution to the overall rate from 
such sources is rather small ($\sim10\%$). This
is especially significant since the calculation for the absolute rate of such
purely hadronic decays are notoriously difficult and it would be virtually
%AS1
impossible to reliably compare the predictions of the SM with experiment
with regard to rates to such a high degree of accuracy.

\vspace*{0.25 in}

We thank  Tom Browder and Jim Smith for useful discussions. This
research was supported in part by the U.S. DOE contracts DC-AC05-84ER40150
(CEBAF), DE-AC-76CH00016 (BNL).

\newpage
\begin{center}
{\large\bf Figure Captions}
\end{center}
\vspace*{.25 in}

Figure 1: (a) The penguin diagram giving rise to $b\to s g^*$ followed by 
$g^*\to g +\eta^\prime$, $f_0$, $f_2$ or other such states.
(b) Contributions to this process from non-standard physics 
%AS1
(NSP) is indicated by the hexagon.

\vspace*{.25 in}

Figure 2: (a) $\pra$ 
%AS1
versus $\mrec$ assuming non-standard physics (NSP) contributes
$\sim 10\%$ to the rate for each state. Also, $\sin\lamF=\sin\lamG=1$ is used.
The black shading shows the $\pra$ for $b\to g+ 0^{-+}$ assuming that $\fx
=0$ and taking $m_{0^{-+}}$ to vary from $958 MeV$ to  $1725 MeV$. The
horizontal striped region is the $\pra$ for the same $0^{-+}$ states, now
assuming that $\gx =0$.  Note that
the region indicated by the diagonal stripes shows the $\pra$ for
$b\to g+ 0^{++}$ assuming that $\fx =0$ with $m_{0^{++}}$ ranging from $980 MeV$
to  $1710 MeV$; the dotted region is the $\pra$ for the $0^{++}$ assuming that
%AS1
$\gx =0$. Note that the PRA for all the $0^{++}$($2^{++}$)  states will have
an opposite sign to that of the $0^{-+}$ states.

(b) $\pra$ versus $\mrec$ assuming NSP contributes $\sim
50\%$ to the rate for each state. The black shading ($\fx =0$) and the
horizontal striped ($\gx =0$) are for $0^{-+}$ states as in fig.~2a. Diagonal
striped ($\fx =0$) and the dotted ($\gx =0$) regions are now for 
$b\to g+2^{++}$
%AS1
with $m_{2^{++}}$ from $1270$ to $2300MeV$. See also caption to fig.~2a.

(c) $\pra$ versus $\mrec$ assuming NSP contributes $\sim 10\%$
to the rate and $G_x=0$. Now the dotted region is for
$0^{-+}$, striped for $0^{++}$ and solid for $2^{++}$.
See also caption to Fig. 2a.

%%%%%%%%%%%%%%%%%%%%%%%%%%%%%%%%%>>>>>>>FIGURE 3 %%%%%%%%%
%                                                        %
%                                                        
\vspace*{.25 in}                                       
Figure 3:  The distribution $d\Gamma/( \Gamma d m_{rec} )$        
in the standard model. 
The solid line is for $J^{PC}=0^{-+}$,                     
the dashed line is for $J^{PC}=0^{++}$, and            
the dotted line is for $J^{PC}=2^{++}$.                
%                                                        %
%                                                        %
%%%%%%%%%%%%%%%%%%%%%%%%%%%%%%%%%%%%%%%%%%%%%%%%%%%%%%%%%%
%
%
%
%
\newpage
\begin{figure}[h]
\epsfysize 6.0in
\mbox{\epsfbox{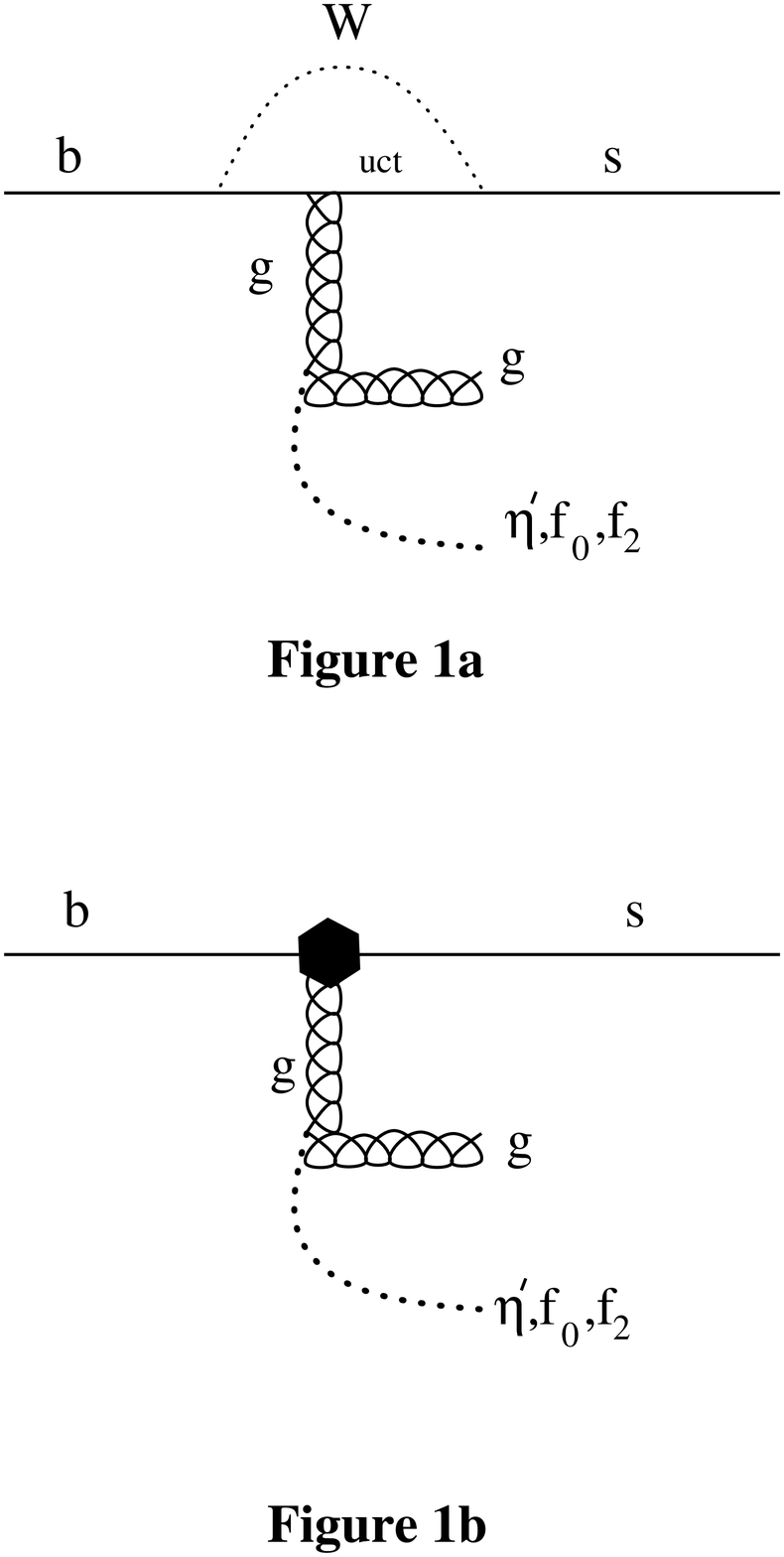}}
\end{figure}
\newpage
\begin{figure}[h]
\epsfysize 6.5in
\mbox{\epsfbox{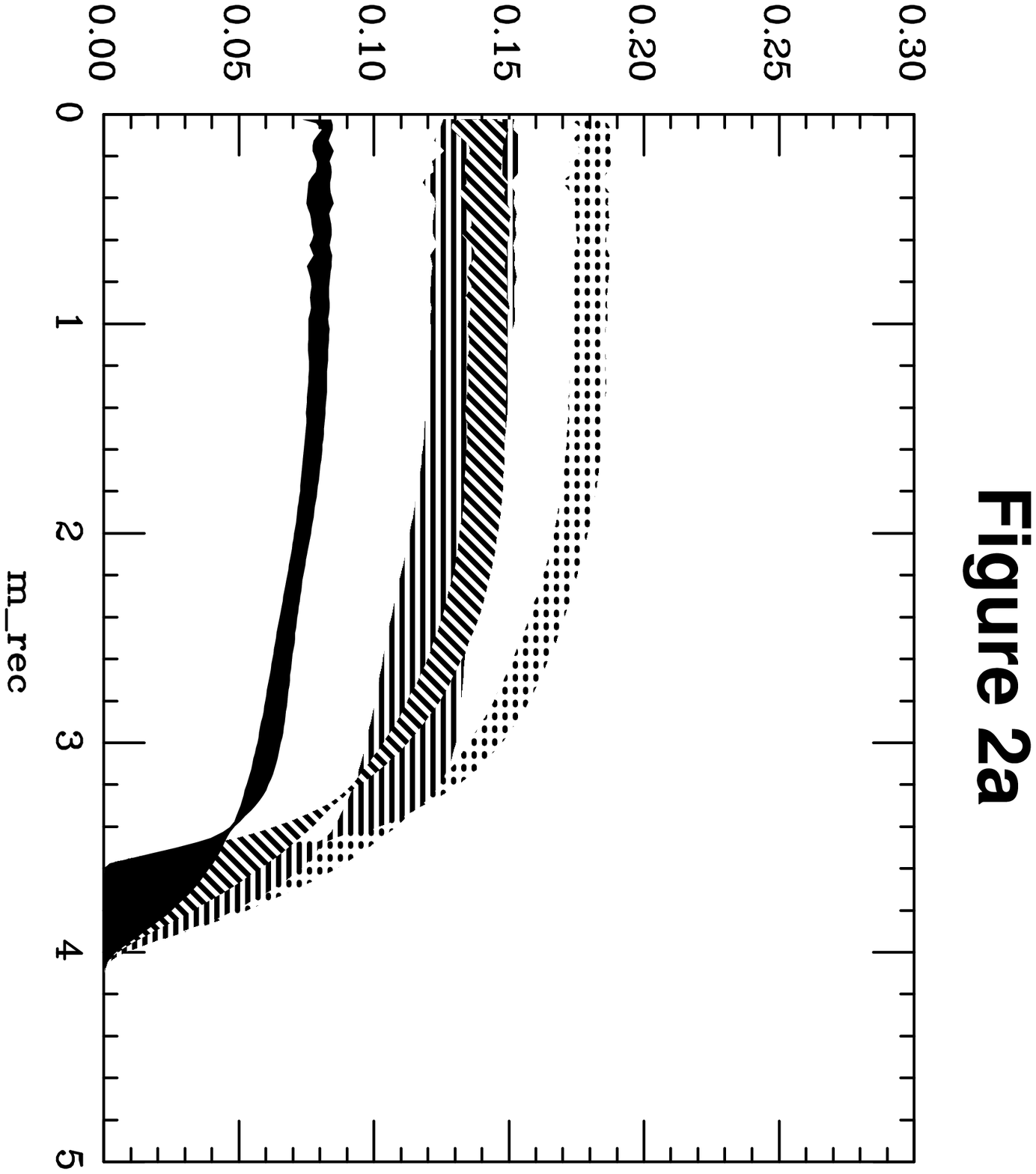}}
\end{figure}
\newpage
\begin{figure}[h]
\epsfysize 6.5in
\mbox{\epsfbox{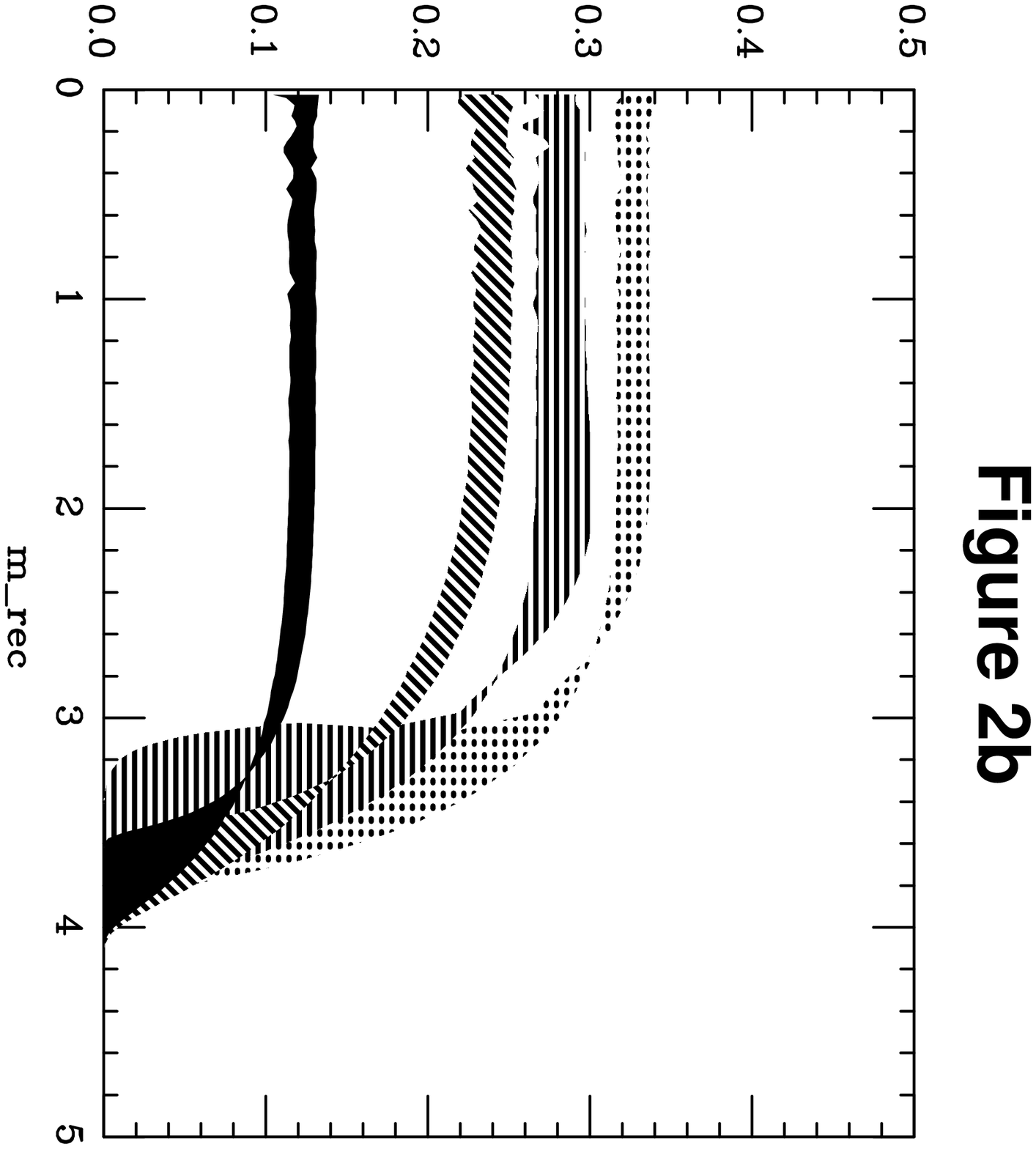}}
\end{figure}
\newpage
\begin{figure}[h]
\epsfysize 6.5in
\mbox{\epsfbox{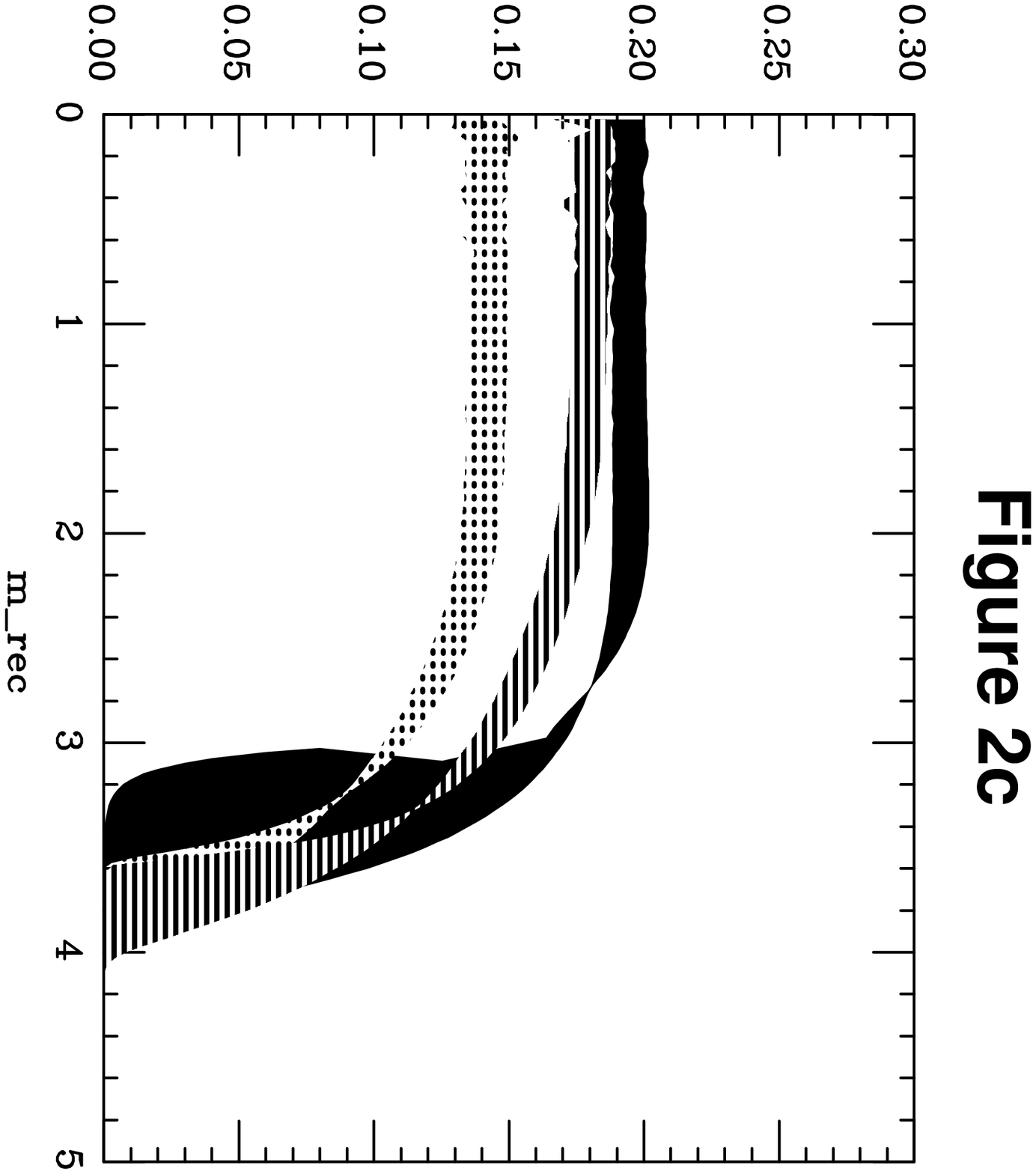}}
\end{figure}
\newpage
\begin{figure}[h]
\epsfysize 6.5in
\mbox{\epsfbox{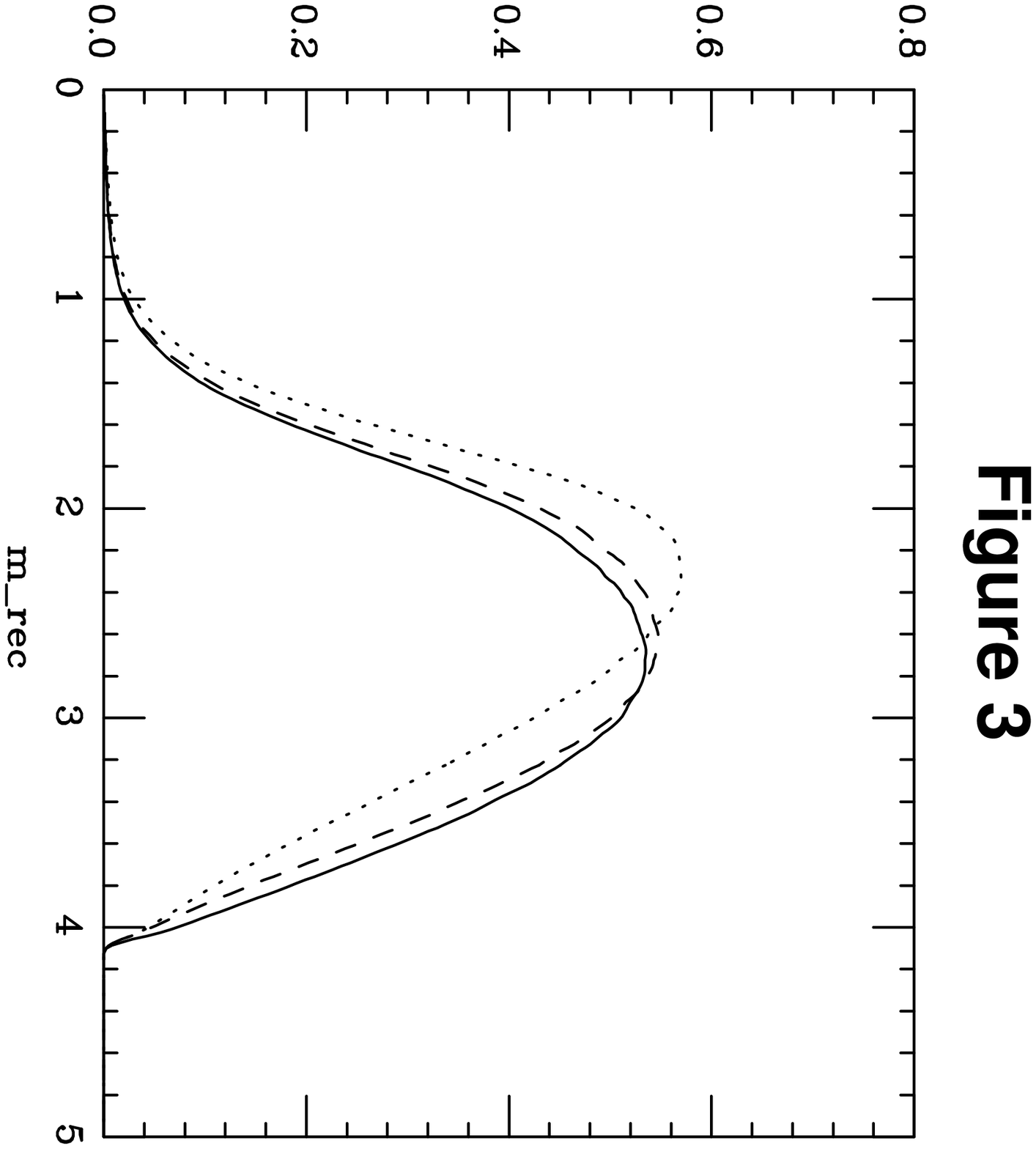}}
\end{figure}

\begin{thebibliography}{99}
\bibitem{ASref_a}
As emphasized in Ref.\cite{ASref} the data suggests that the $g^*-g-
\eta^\prime$ vertex is being influenced by the presence of near-by
resonances, as it should become even more apparent in this Letter.
Thus the form-factor tends to stay fairly constant even for rather large
$q^2$. After that key assumption is made, which requires non-perturbative
physics  to modify a power-like $q^2$ dependence, perturbative effects
(such as running of $\alpha_s$, which can only have a logarithmic
%AS1
dependence on $q^2$) cannot be applied reliably. Thus in Ref.\cite{ASref},
QCD corrections  were not taken into account seriously for the
chromo-electric form factor, rather they were primarily considered to 
include their effects only on the chromo-magnetic form-factor (which is non-leading
anyway) as it is rather sensitive to radiative effects.
\bibitem{glueballs}
Although these states are expected to be rich in their gluonic
content, they are not necessarily glueballs.
\bibitem{hel_note}
We are assuming here that non-standard physics only modifies the form-factors
(F and G) and not the chirality (L and R) to enable the new physics to interfere
with the SM without entailing a $m_s/m_b$ suppression.
\bibitem{eta*note}
The $\eta(1440)$ deserves special experimental attention. Inspection of $\psi$
decays into final states containing
$\eta^\prime(958)$ or $\eta(1440)$ suggests that the 
$Br(B\to X_s\eta(1440))$ is likely to be in the same ball-park as 
$Br(B\to X_s\eta^\prime(958))$.
\bibitem{vector_note}
%AS1
Although, in perturbation theory, one expects the production of  $1^{--}$
states  through our fragmentation mechanism to be suppressed as
they require an extra gluon, some of the vector states may well be more readily
accessible experimentally  to offset such a supression. Notice, in particular,
$\phi$ which has a distinctive signature ($\phi\to K\bar K$) 
falls into this class of interesting examples of $X_g$.
\end{thebibliography}
\end{document}